\documentclass[aps,prc,twocolumn,floatfix,12pts,superscriptaddress]{revtex4}
\usepackage{epsfig}
\usepackage{amssymb}
\usepackage{amsmath}
\usepackage{color}
\usepackage{graphicx}
\usepackage{epsfig}
\usepackage{amsmath}
\usepackage{color}
\usepackage{graphicx}
\usepackage{multirow}
\usepackage{hyperref}
\usepackage{ulem}
\usepackage{float}
\definecolor{blue}{rgb}{0.05, 0.05, 0.5}
\def \beq{\begin{equation}}
\def \eeq{\end{equation}}
\def \beqa{\begin{eqnarray}}
\def \eeqa{\end{eqnarray}}

\usepackage{xspace}

\begin{document}
\title{The neutron skin effect in Pb+Pb collisions at 2.76A TeV at the LHC}
\author{Amit Paul}
\email{a.paul@vecc.gov.in}

\author{Rupa Chatterjee}
\email{rupa@vecc.gov.in}
\affiliation{Variable Energy Cyclotron Centre, 1/AF, Bidhan Nagar, Kolkata-700064, India}
\affiliation{Homi Bhabha National Institute, Training School Complex, Anushaktinagar, Mumbai 400094, India}

\begin{abstract}
Collisions of lead nuclei at relativistic energies provide valuable insight into the properties of the quark gluon plasma formed in such collisions where the initial geometry and density profile play a crucial role in governing the subsequent evolution of the produced hot and dense fireball.
The neutron skin thickness resulting from the difference between the neutron and proton density distributions in neutron rich lead nuclei plays an important role in  nuclear structure studies.  In this work we investigate the impact of  neutron skin on the space time evolution of the fireball formed in Pb+Pb collisions at 2.76A TeV at the LHC and analyze how the presence of  neutron skin affect bulk observables sensitive to the initial nuclear structure. The time evolution of initial profile along with the average $p_T$, particle spectra and anisotropic flow parameters are estimated to investigate the effect of neutron skin on these observables. The initial spatial anisotropy of the fireball is found to be affected by the neutron skin thickness significantly especially for the peripheral collisions. This leads to a substantial enhancement of the elliptic flow of hadrons with an even stronger effect observed for photons. In addition, the effect is found to be more pronounced for lower beam energy collisions of lead nuclei.

\end{abstract}

\maketitle

\section{Introduction}
In neutron rich nuclei the differences in the spatial distributions of protons and neutrons result in the formation of a neutron skin~\cite{prl,Tarbert:2014, prl1}. The neutron skin thickness  defined as the difference between the root mean square radii of the neutron and proton distributions quantifies this effect. The equation of state of the nuclear matter determines properties  starting from atomic nuclei to neutron stars and the effects of neutron proton asymmetry is characterized by its symmetry energy. Thus,  precise estimation of the neutron skin thickness offer important constraints on the nuclear symmetry energy~\cite{s0,s1,s2,Nijs:2023, s3, s4}.  

The measurement of the neutron skin thickness through parity violating electron scattering by the PREX Collaboration provided the  first direct observation of it in a heavy neutron rich nucleus~\cite{prex}. This experiment determined  the difference between the proton and neutron radii thereby quantifying the neutron skin thickness in doubly magic $^{208}$Pb (Z = 82, N = 126).  The reported  neutron skin thickness by the PREX experiment is $\Delta_{np} \approx 0.28\,\mathrm{fm}$ for ${}^{208}\mathrm{Pb}$.  Recent studies have shown that the presence of neutron skin thickness in neutron rich heavy nucleus can measurably influence observables in relativistic nuclear collisions~\cite{De:2016ggl, s7, Paukkunen:2015, Helenius:2017, Li:2020, Schee:2024, Pihan:2025, J.Hammelmann:2020, song}.

Collisions of lead nuclei at top LHC energies have provided strong evidence for the formation of hot and dense quark gluon plasma (QGP) in these collisions and the measured observables have yielded valuable information on the properties of this strongly interacting medium. Detailed analyses of a wide range of bulk and differential observables including charged particle multiplicity, particle spectra, anisotropic flow coefficients, electromagnetic probes, heavy flavor/quarkonia have significantly advanced our understanding of the bulk and transport properties of the QGP as well as the dynamics of its space time evolution~\cite{qgp1, qgp2, qgp3, qgp4, v2_1, v2_2}.


 In neutron rich nuclei where the neutron distribution is more extended than the proton distribution, the initial state is constructed using separate neutron and proton density profiles instead of a single nucleon distribution~\cite{Paukkunen:2015}.
The presence of a neutron skin  in lead nuclei  thus modifies the (single) nucleon density distribution and consequently influence the observables measured in collisions of Pb+Pb at relativistic energies. Relativistic hydrodynamics is one of the most powerful frameworks to model the evolution of the hot and dense QGP matter produced and to study the bulk observables~\cite{kolb, B.Schenke:2010}. The initial state of the model calculation depends on the initial nuclear density distribution.  Thus, a modification in the initial density distribution is expected to affect the evolution of the QGP and hot hadronic matter along with various final state observables estimated  using that.

We study the effect of neutron skin  on observables measured in Pb+Pb collisions at 2.76A TeV at the LHC using a hydrodynamical framework employing Glauber model initial conditions. The effect on initial spatial anisotropy and evolutions of the fireball are studied along with hadron spectra, average $p_T$ and anisotropic flow of hadrons considering with and without neutron skin effect. 

Thermal photons are emitted throughout the evolution of the hot and dense medium and thus, they are  expected to be more sensitive to the neutron skin effect (NSE) compared to hadrons which are  emitted from the surface of freeze-out~\cite{dks, gale, gabor}. The $p_T$ spectra and anisotropic flow of thermal photons are also evaluated by incorporating the neutron skin. The results are compared with those obtained using an initial state based on a single nucleon density distribution without incorporating the neutron skin. Our results indicate a significant influence of the neutron skin on the system evolution and on anisotropic flow coefficients which should be included in model calculations when comparing with experimental data.

\section{Framework}
 In the typical Glauber model formalism, the nucleon density distribution is described by a single parameterized profile with a common nuclear radius and skin thickness without distinguishing between proton and neutron distributions.

For the nucleon density $\rho_A$ in a spherical nucleus of mass number $A$ we use the two parameter Woods–Saxon distribution:
\begin{equation}
\rho_A(\mathbf{r}) = \frac{\rho_{0,A}} {1 + \exp\left[\frac{|\mathbf{r}| - d_A}{a_A}\right]} 
\end{equation}
where, $a_A$ the nuclear skin thickness and $d_A$ is the nuclear radius. For lead, the nuclear radius $d_{\text{Pb}} = 6.624~\mathrm{fm}$ and nuclear skin thickness $a_{\text{Pb}} = 0.549~\mathrm{fm}$~\cite{kolb}.

This initial state with single nucleon density distribution without the neutron skin effect here after will be denoted as w/o NSE case.

The PREX experiment results indicate that the neutron distribution differs from the proton distribution resulting in finite neutron skin thickness. In that case, the total nuclear density is  expressed as a combination of the individual proton and neutron density distributions characterized by different radii and surface diffuseness parameters in the lead nucleus.

We refer to the initial state constructed using two separate density distributions as the with neutron skin effect (w NSE) case~\cite{Tarbert:2014, Paukkunen:2015,Helenius:2017}:
\begin{equation}
    \rho_A(\mathbf{r})=\rho_A^p(\mathbf{r})+\rho_A^n(\mathbf{r})
\end{equation}
where,
\begin{equation}
\rho_A^i(\mathbf{r}) = \frac{\rho_{0,A}^i} {1 + \exp\left[\frac{|\mathbf{r}| - d_A^i}{a_A^i}\right]} 
\end{equation}
Here, $i = p, n$ in the above Woods-Saxon density distribution.

We have considered $d_{\text{Pb}}^p$ = 6.680 fm and $a_{\text{Pb}}^p$ = 0.447 fm for protons and $d_{\text{Pb}}^n$ = 6.70 ± 0.03 fm and $a_{\text{Pb}}^n$ = 0.55 ± 0.03 fm for neutrons~\cite{Tarbert:2014,Paukkunen:2015,Helenius:2017}. 

The saturation densities $\rho_{0,A}^i$ are obtained by conserving the total number of protons and neutrons in lead nucleus separately,
\begin{align}
\int d^3\mathbf{r} \, \rho_{A}^p(\mathbf{r}) &= Z = 82,\\
\int d^3\mathbf{r} \, \rho_{A}^n(\mathbf{r}) &= N = 126.
\end{align}
\begin{figure}
    \centerline{\includegraphics*[scale=0.32,clip=true]{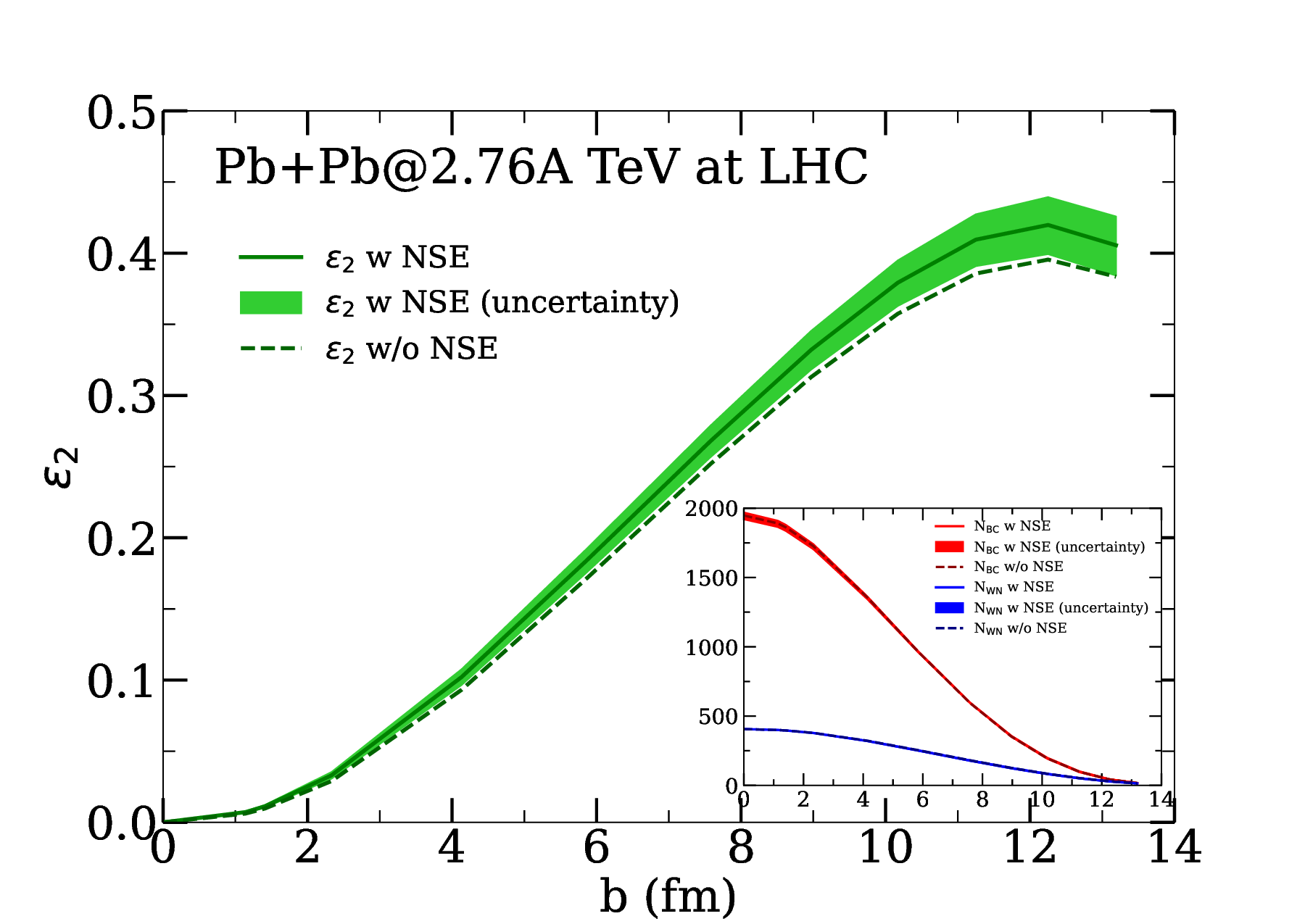}}
	\caption{(Color online) Centrality dependent initial spatial anisotropy ($\varepsilon_2$) and  $N_{\rm WN}$ and $N_{\rm BC}$ [inset] in Pb+Pb collisions at 2.76A TeV at the LHC with and without neutron skin effect.}
	\label{fig_npart}
\end{figure}
\begin{figure}
	\centerline{\includegraphics*[scale=0.35,clip=true]{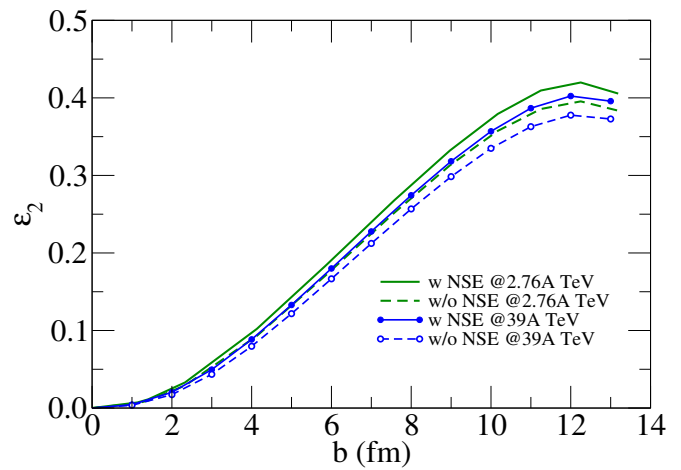}}
	\caption{(Color online) Centrality dependent initial spatial anisotropy $\varepsilon_2$ at 2.76 and 39 A TeV Pb+Pb collisions at the LHC and FCC respectively with and without neutron skin effect.}
	\label{fig.fcc}
\end{figure}

In Glauber model formalism with smooth initial density distribution,  the wounded nucleons per transverse area in collision of nuclei $A$ and $B$  at an impact parameter $b$ is defined as~\cite{kolb, M.Miller:2007}: 
\begin{align}
n_{\text{WN}}(x,y;b) =
&\, T_A\!\left(\mathbf{s}_1\right)
\left[
1 - \left(
1 - \frac{\sigma^{\text{inel}} T_B\!\left(\mathbf{s}_2\right)}{B}
\right)^{B}
\right] \nonumber \\
&\quad + T_B\!\left(\mathbf{s}_2\right)
\left[
1 - \left(
1 - \frac{\sigma^{\text{inel}} T_A\!\left(\mathbf{s}_1\right)}{A}
\right)^{A}
\right]  \ .
\end{align}

 Whereas, the binary nucleon nucleon collisions per transverse area is defined as: 

\begin{align}
n_{\text{BC}}(x,y; b) =\sigma^{\text{inel}}
T_A\!\left(\mathbf{s}_1\right)
T_B\!\left(\mathbf{s}_2\right) 
\end{align} 
where, $\sigma^{\text{inel}}$ is the nucleon–nucleon inelastic cross section and is taken as 64 mb for Pb+Pb collision at 2.76A TeV. $T_A\left(\mathbf{s}_1\right)$ and $T_B\left(\mathbf{s}_2\right)$ are the corresponding nuclear thickness functions for nuclei $A$ and $B$ respectively and $\mathbf{s}_{1,2} = (x \pm b/2, y)$.

For the NSE, the thickness functions are given by:
\begin{align}
T_A\left(\mathbf{s}_1\right) &= T_A^p\!\left(\mathbf{s}_1\right) + T_A^n\!\left(\mathbf{s}_1\right)\\
T_B\left(\mathbf{s}_2\right) &= T_B^p\!\left(\mathbf{s}_2\right) + T_B^n\!\left(\mathbf{s}_2\right)
\end{align}
where, $T_A^i(\mathbf{s})$ are calculated by integrating the density of nucleons $i$ over the longitudinal direction:
\begin{equation}
T_A^i(\mathbf{s}) = \int dz \, \rho_A^i(\mathbf{r}) \ .
\end{equation}

A longitudinally boost invariant (2+1) dimensional ideal hydrodynamical model (MUSIC)~\cite{B.Schenke:2010} has been used to study the evolution of the hot and dense matter produced in 2.76A TeV Pb+Pb collisions at the LHC considering with and without neutron skin effect. The initial energy density distribution (at $\tau_0$) on the transverse plane $(x,y)$ is taken as a linear combination as

\begin{equation}
\varepsilon(x, y; b) = K \left[ (1 - \alpha)\, n_{\text{WN}}(x, y; b) + \alpha\, n_{\text{BC}}(x, y; b) \right]
\end{equation}
where, the factor $\alpha$ is taken as 0.05. The constant $K$ in equation above and the initial parameters are adjusted to reproduce the experimental data for charged particle spectra and multiplicity for Pb+Pb collisions at 2.76A TeV. 


The spatial eccentricity $\varepsilon_2$ is obtained from the relation:
\begin{equation}
    \varepsilon_2(b, \tau) = \frac{\langle y^2 - x^2 \rangle}{\langle y^2 + x^2 \rangle}.
\label{ecc}
\end{equation}
\begin{figure}
	\centerline{\includegraphics*[scale=0.35,clip=true]{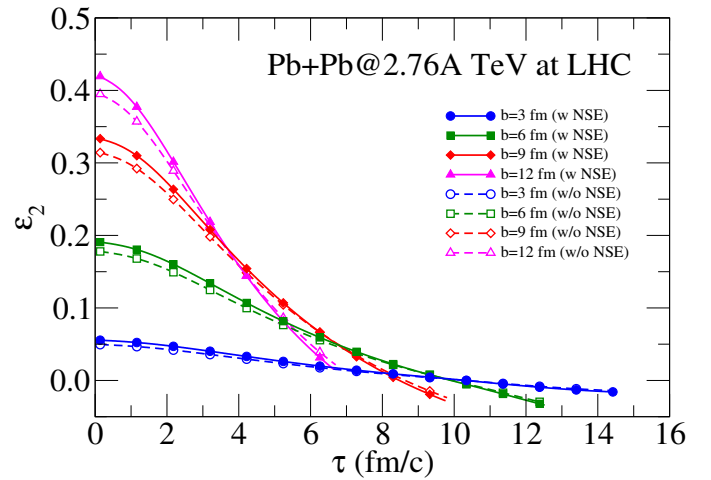}}

	\caption{(Color online) Time evolution of spatial eccentricity  $\varepsilon_2$ at different impact parameters of 2.76A TeV Pb+Pb collisions at the LHC with and without neutron skin effect.}
	\label{fig.2}
\end{figure}

An initial formation time of $\tau_0 = 0.14~\text{fm}/c$~\cite{tau_2} is assumed and the corresponding central energy density is taken to be 
$576~\text{GeV}/\text{fm}^3$ for Pb+Pb collisions at 2.76A TeV.  A lattice based equation of state is considered~\cite{P.Huovinen:2010} with a final freeze out temperature as 145 MeV. The total charged particle multiplicity is estimated to be about 1600 for the most central collisions at 2.76A TeV at LHC using these set of initial parameters. The calculations are extended for 39A TeV Pb+Pb collisions at FCC considering same  $\tau_0$ value of  0.14 fm/$c$, freeze-out temperature as 145 MeV and charged particle multiplicity of about 3600~\cite{Dasgupta:2018pjm}.

The Cooper-Frye formalism is used for the hadron production from freeze-out surface~\cite{Cooper:1974}. Whereas, for thermal photons the thermal emission rates from quark matter and hadronic matter phases are integrated over the entire space time history. The standard thermal photons rates are taken from~\cite{Arnold:2001ms,Ghiglieri:2013gia, Turbide:2003si}.

\section{Results and discussions}

\begin{figure}
	\centerline{\includegraphics*[scale=0.4,clip=true]{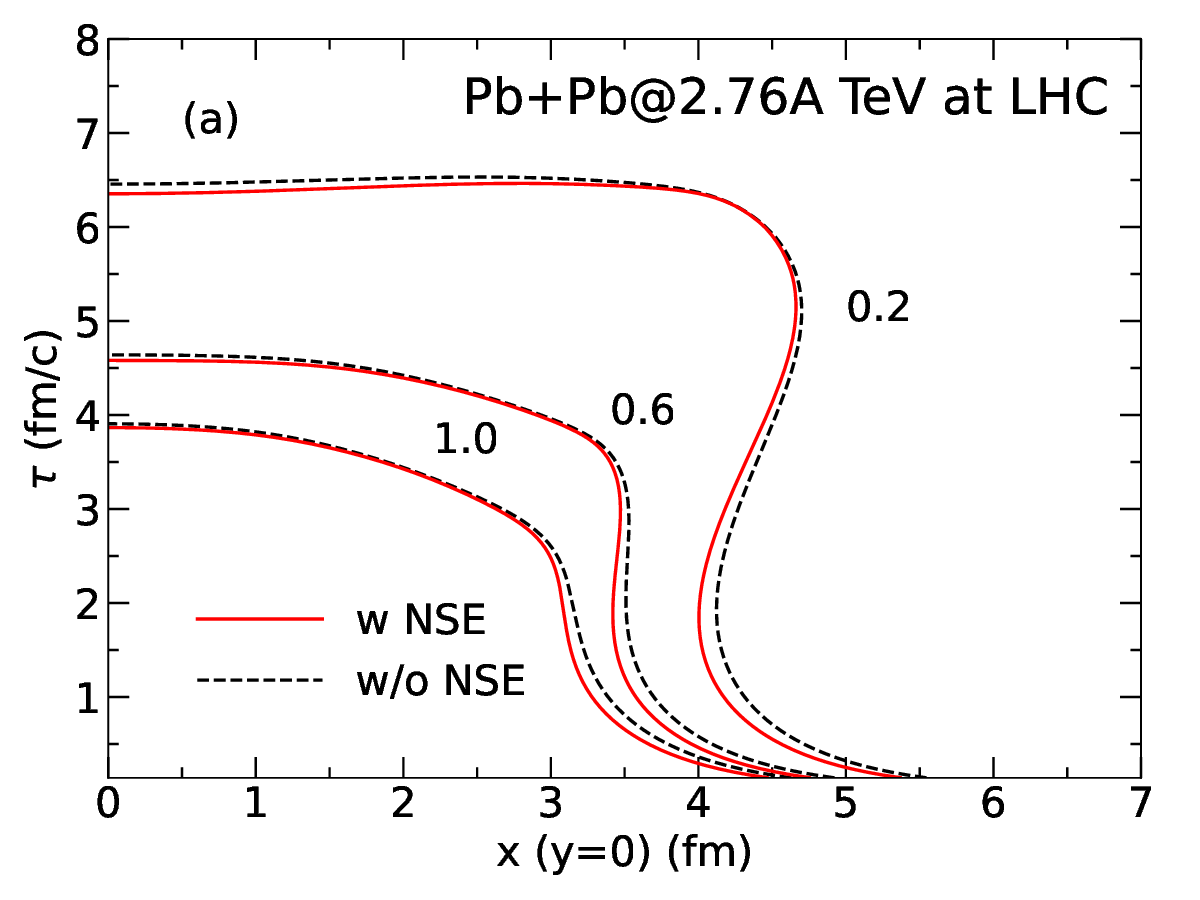}}
    \centerline{\includegraphics*[scale=0.4,clip=true]{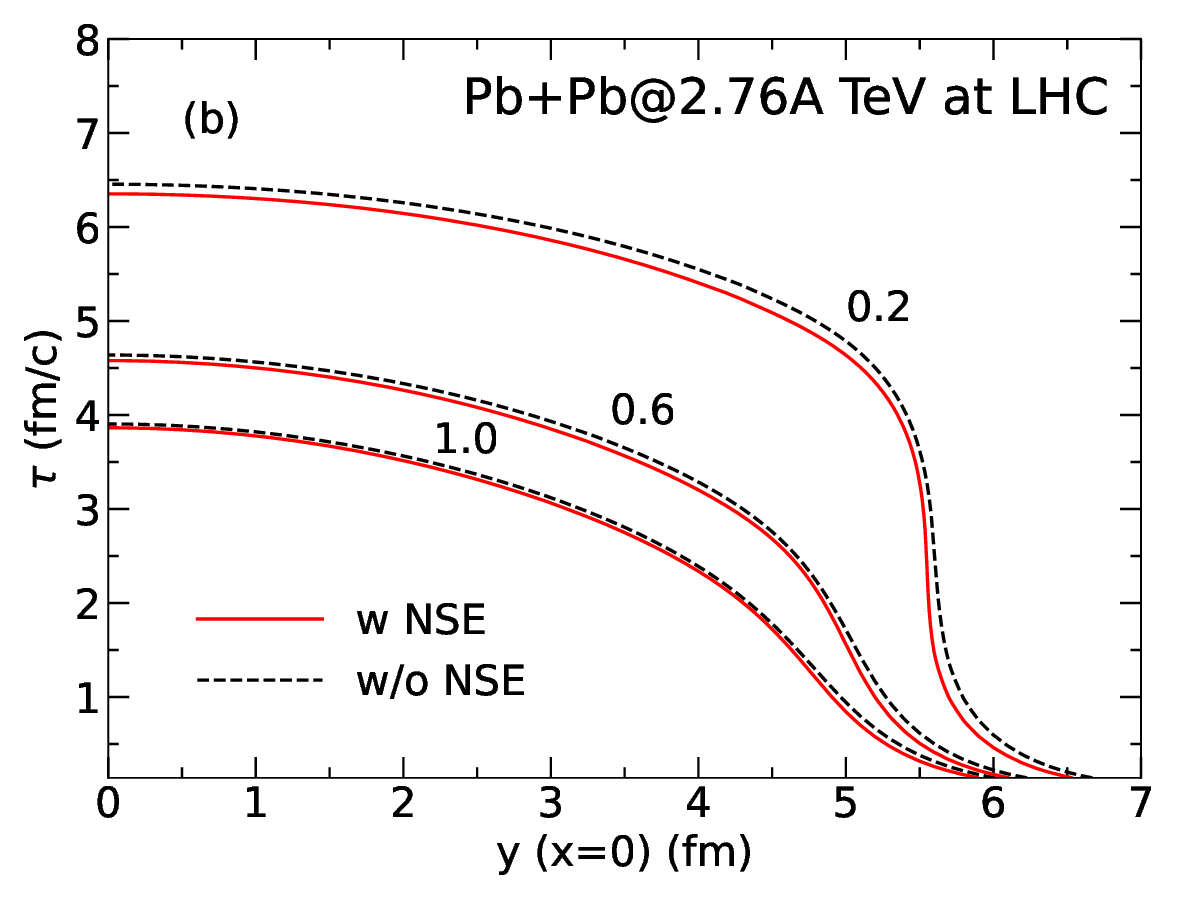}}

	\caption{Constant energy density contours (in GeV) along the (a) x direction (at y=0) and (b) y direction (at x=0) in Pb+Pb collisions at 2.76A TeV at the LHC for impact parameter b=12 fm. Results are shown for initial conditions with and without neutron skin.}
	\label{cont}
\end{figure}


The initial spatial eccentricity $\varepsilon_2$ along with number of participants  $N_{\rm WN}$ and the number of binary collisions $N_{\rm BC}$ for Pb+Pb collisions at 2.76A TeV are shown  in Fig.~\ref{fig_npart} considering with and without NSE. 

The total number of participants and  binary collisions  at a particular impact parameter are estimated by integrating  $n_{\rm WN} (x,y,b)$ and $n_{\rm BC}(x,y,b)$ over the transverse plane.  The maximum uncertainty in the results arising from the experimental uncertainty in the estimation of neutron skin thickness is represented by the uncertainty band in the figure. Both  $N_{\rm WN}$ and  $N_{\rm BC}$ shows only marginal variations due to the inclusion of neutron skin thickness as illustrated in the inset of Fig.~\ref{fig_npart}.

In contrast, the initial spatial eccentricity $\varepsilon_2$ is found to be significantly influenced by the neutron skin thickness. This effect is more pronounced in peripheral collisions compared to central collisions indicating that observables sensitive to neutron skin effects would exhibit a stronger response in peripheral events.  

A comparison showing the impact of the neutron skin on the initial eccentricity $\varepsilon_2$ at a significantly higher beam energy of 39A TeV is presented in Fig.~\ref{fig.fcc}. The centrality dependence of the difference between the two cases is more pronounced for Pb+Pb collisions at 2.76A TeV at the LHC than for Pb+Pb collisions at 39A TeV at the FCC. For the FCC energy the inelastic nucleon–nucleon cross section is taken to be 80 mb \cite{Dasgupta:2018pjm}.

The variation of the spatial eccentricity $\varepsilon_2$ as a function of proper time $\tau$ is shown in Fig.~\ref{fig.2}. Eccentricity values are shown at different impact parameters to compare the relative impact of NSE on the time evolution of $\varepsilon_2$. The difference in eccentricity values between with and without NSE is observed to be largest at the early stages of the evolution as shown in the figure.

\begin{figure}
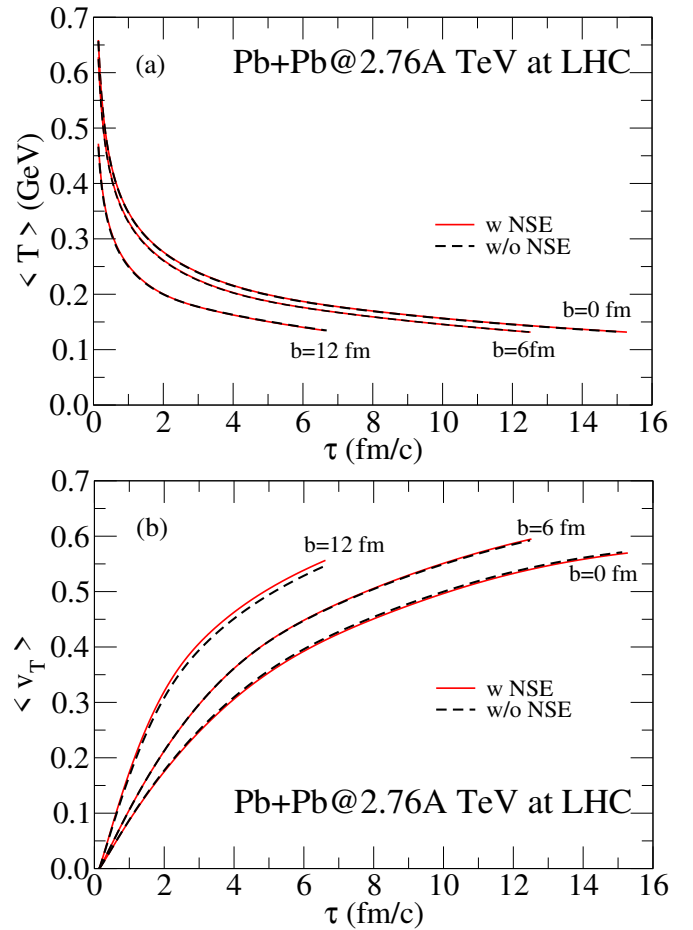

	\centerline{\includegraphics*[scale=0.35,clip=true]{temp.eps}}
    \centerline{\includegraphics*[scale=0.35,clip=true]{vt.eps}}

	\caption{(Color online) Time evolution of (a) the average temperature $\langle T \rangle$ and (b) the average transverse flow velocity $\langle v_T \rangle$ from Pb+Pb collisions at 2.76A TeV at LHC for different impact parameters. Results are shown for initial conditions with and without neutron skin.}
	\label{evol}
\end{figure}

The constant energy density contours illustrate how the space time evolution of the system is modified by the inclusion of neutron skin effects in the initial state. Fig.~\ref{cont}  shows the energy density profiles along the $x$ direction at $y=0$ and along the  $y$ direction at $x=0$  respectively for an impact parameter $b=12$ fm. The contours are shown for energy density values 1.0, 0.6 and 0.2 GeV/fm$^3$.

As the system evolves with time the expansion of the energy density contours is found to be slightly more in the case without neutron skin compared to the case where neutron skin is included. This indicates a small variation of the transverse expansion dynamics due to neutron skin effect. The difference between the two cases becomes more noticeable near the outer boundary of the system, where the initial geometry and surface structure play a more significant role in the subsequent evolution.

Next, we estimate the time evolution of the average temperature $\langle T \rangle$ and 
 and the average transverse flow velocity $\langle v_T \rangle$ for different impact parameters as shown in Fig.~\ref{evol}. 
 The effect of neutron skin on the average temperature is  found to be only marginal for both central and peripheral collisions of Pb+Pb at 2.76A TeV.  However, the $\langle v_T \rangle$ results obtained by including neutron skin exhibit a slightly larger buildup compared to the case without neutron skin. 

\begin{figure}
	\centerline{\includegraphics*[scale=0.35,clip=true]{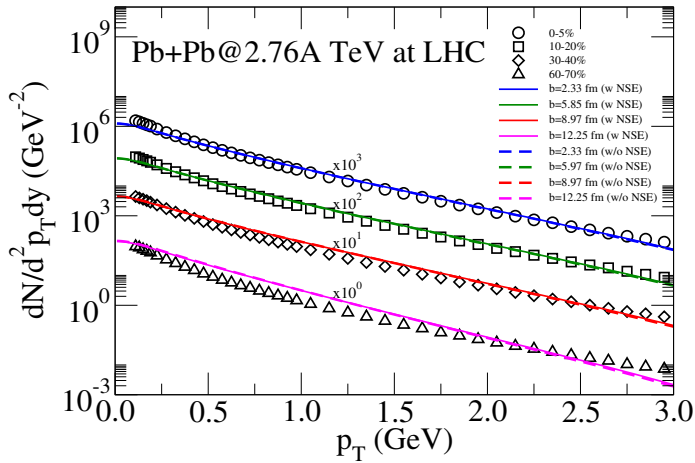}}
    
	\caption{(Color online) Charged pion spectra calculated with and without neutron skin compared with ALICE data for different centrality bins from Pb+Pb collisions at 2.76A TeV at the LHC~\cite{Abelev:2013}.}
	\label{fig.5a}
\end{figure}

\begin{figure}
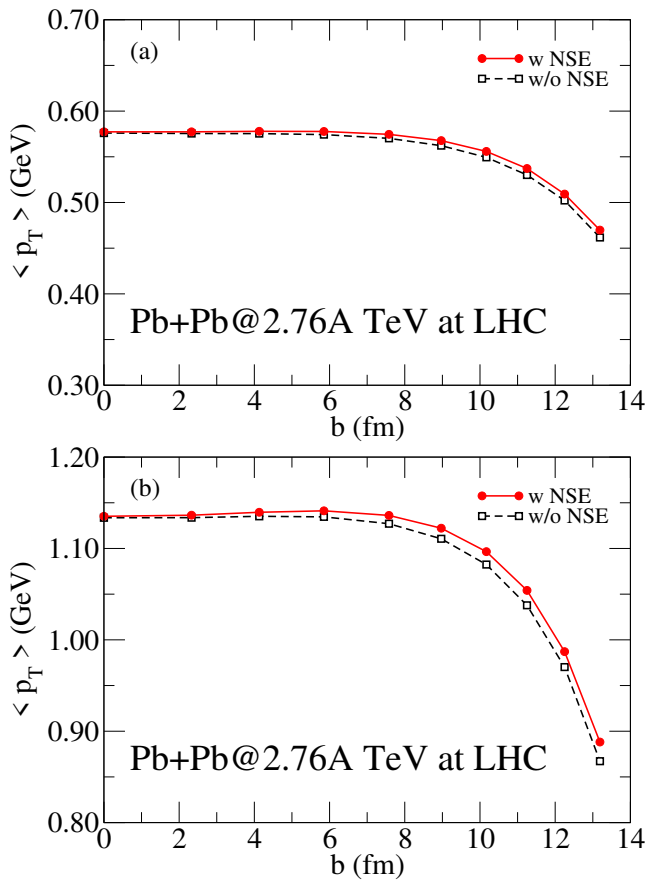

	\centerline{\includegraphics*[scale=0.33,clip=true]{pt_pi.eps}}
    \centerline{\includegraphics*[scale=0.33,clip=true]{pt_p.eps}}

	\caption{(Color online) Average $p_T$ of (a) charged pion  and (b) proton as a function of impact parameter for 2.76A TeV Pb+Pb collisions at LHC with and without neutron skin effect.}
	\label{fig_pt}
\end{figure}

The charged pion spectra calculated within the hydrodynamical model framework are found to describe the experimental data well for Pb+Pb collisions at 2.76A TeV at the LHC across different centrality classes [Fig.~\ref{fig.5a}]. The impact parameters b=2.33, 5.85, 8.97, and 12.25 fm correspond to the centrality bins 0–-5\%, 10–-20\%, 30–-40\%, and 60–-70\% respectively. The inclusion of neutron skin effect is found to have a negligible influence on the charged particle spectra. Consequently, the spectra obtained with and without NSE are found to be similar and both describe the experimental data well up to a large value of $p_T$.

The effect of neutron skin on the average transverse momentum $\langle p_T \rangle$ of charged particles as a function of collision centrality is shown in Fig.~\ref{fig_pt}. The  $\langle p_T \rangle$ reflects the buildup of collective flow driven by pressure gradients in the medium. A stronger collective expansion leads to a larger values of $\langle p_T \rangle$. The average transverse momentum also exhibits mass ordering a key signature of hydrodynamic expansion with heavier particles showing larger $\langle p_T \rangle$.

 The inclusion of neutron skin in the Glauber model initial conditions results in a slight enhancement of $\langle p_T \rangle$ for charged pions and protons as shown in the figure. This enhancement is found to be slightly more pronounced in peripheral collisions compared to central collisions.
\begin{figure}
	\centerline{\includegraphics*[scale=0.35,clip=true]{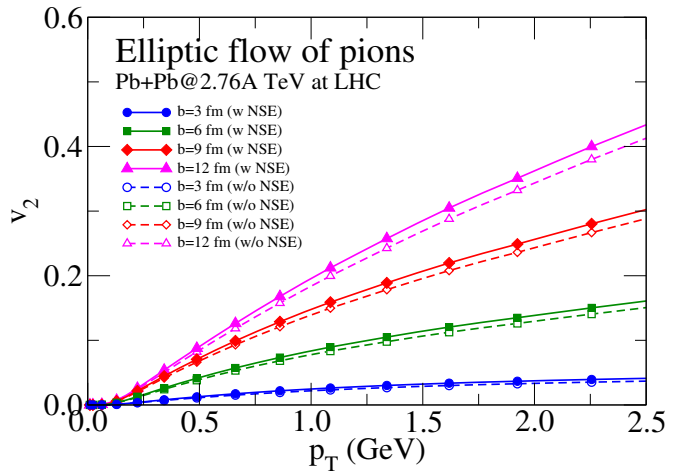}}
    
	\caption{(Color online) Elliptic flow of charged pions from 2.76A TeV Pb+Pb collisions at LHC at different values of impact parameter with and without neutron skin effect.}
	\label{pi_v2}
\end{figure}

Although the transverse momentum spectra of charged pions are only marginally affected by neutron skin effects, a pronounced sensitivity is observed in the anisotropic flow coefficients as shown in Fig.~\ref{pi_v2}. The difference in the initial spatial eccentricity due to NSE is reflected in the pion anisotropic flow with the pion $v_2$ as a function of $p_T$ for peripheral collisions  being larger (maximum difference at b=12 fm) when neutron skin is included in the initial conditions.

Earlier studies have shown that the presence of neutron skin can influence prompt photon production in relativistic nuclear collisions substantially~\cite{Helenius:2017, De:2016ggl}. Thermal photons on the other hand are emitted continuously throughout the entire space time evolution of the hot and dense medium. As a result, neutron skin effects are expected to have a stronger impact on thermal photons than prompt and hadronic observables since they carry information from all stages of the collision without undergoing strong final state interactions.

\begin{figure}
	\centerline{\includegraphics*[scale=0.35,clip=true]{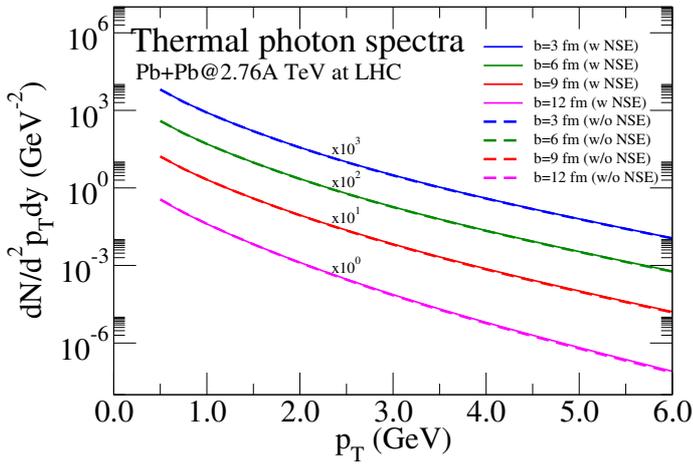}}
    
	\caption{(Color online) Thermal photon $p_T$ spectra at different impact parameters from 2.76A TeV Pb+Pb collisions at the LHC with and without neutron skin effect. }
	\label{phot_spec}
\end{figure}

\begin{figure}
	\centerline{\includegraphics*[scale=0.35,clip=true]{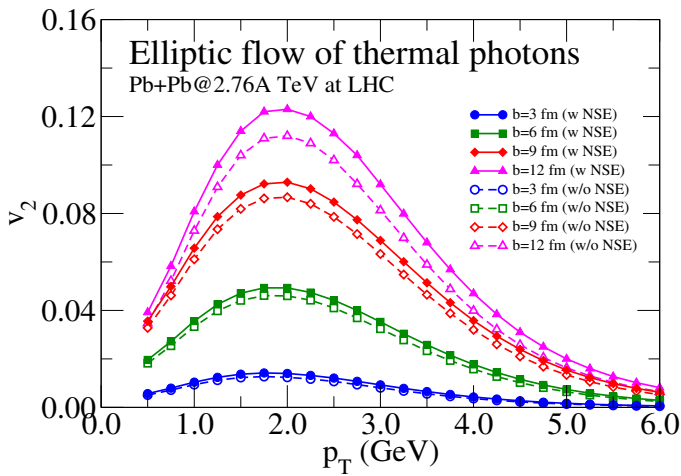}}
    
	\caption{(Color online) Elliptic flow of thermal photons $v_2(p_T)$ from Pb+Pb collisions at 2.76A TeV at the LHC at different impact parameters with and without neutron skin effect.}
	\label{phot_v2}
\end{figure}

The thermal photon transverse momentum spectra however, exhibit only a weak sensitivity to neutron skin effects as shown in Fig.~\ref{phot_spec} similar to the behavior  observed for pion spectra. In contrast,  the anisotropic flow of thermal photons shows much stronger sensitivity to the neutron skin as shown in Fig.~\ref{phot_v2}. This is due to the modification of the initial spatial eccentricity and its subsequent influence on the development of collective flow which is more directly reflected in photon anisotropic flow observables~\cite{rc_prl, gale1, rc_1}.

These results indicate that neutron skin thickness plays  an important role in the estimation of anisotropic flow parameters of photons and hadrons in model calculations and should be taken into account when comparing with experimental data.

\section{Summary and conclusions}
The finite neutron skin thickness in neutron rich lead nuclei arises from the different spatial distributions of neutrons and protons as established by nuclear structure studies. The impact of this neutron skin on the fireball evolution and on the resulting observables in relativistic Pb+Pb collisions at 2.76 TeV at the LHC has been investigated using a relativistic ideal hydrodynamic framework. 

 The Glauber model initial conditions show that incorporating neutron skin thickness through individual proton and neutron density distributions results in only marginal changes in the number of participating nucleons and binary collisions compared to results obtained using a single nucleon  distribution. In contrast, the spatial eccentricity $\varepsilon_2$ is found to be significantly affected by the inclusion of the neutron skin, particularly for peripheral collisions and during the first few fm/$c$ time period of the fireball evolution. The value of $\varepsilon_2$ is found to increase systematically toward more peripheral collisions when neutron skin effect is included.

At 39A TeV Pb+Pb collisions, a smaller relative increase in spatial eccentricity is observed indicating reduced sensitivity to neutron skin effect at higher beam energies.

The hydrodynamic evolution of the hot and dense medium further shows that the time dependence of the average temperature is essentially insensitive to the presence of the neutron skin. However, a mild modification of the average transverse flow velocity is observed when NSE are taken into account especially in peripheral collisions. The medium expansion is also influenced by NSE as reflected in the differences in constant energy density contours obtained with and without the neutron skin effect.

In addition, the average transverse momentum $\langle p_T \rangle$ of charged pions and protons is found to be slightly larger when NSE is considered. While the pion and thermal photon $p_T$ spectra remain largely unaffected, the elliptic flow coefficient exhibits significant sensitivity to the neutron skin. This effect is found to be  more pronounced for thermal photons with a substantial enhancement of elliptic flow observed in peripheral collisions when NSE is included in the calculation.

Overall, these results show that neutron skin effect can play a significant role in shaping the medium evolution and final state observables in relativistic Pb+Pb collisions particularly in peripheral events. While the present theoretical model calculations of photon anisotropic flow continue to underestimate the corresponding experimental measurements, an issue that remains an important open problem in photon studies of heavy ion collisions~\cite{ratio, ratio1}. 

Our initial results indicate that the inclusion of NSE can partially reduce the discrepancy between  results from theoretical model calculation and the experimental  data for peripheral collisions. A more detailed analysis incorporating event-by-event fluctuations and viscous effects is required for a quantitative assessment of neutron skin contributions. However, the qualitative and  generic nature of the conclusions presented in this work are expected to remain robust.



\begin{acknowledgments}
We would like to thank Somnath De for useful discussions. A.P. acknowledges the computing facility KANAAD of the VECC Computer Centre.
\end{acknowledgments}

%

%

\end{document}